\documentclass[12pt]{article}
\usepackage{a4wide,amsmath,amssymb,graphicx,psfrag,feynarts}

\newcommand{\E}[1]{\cdot 10^{#1}}

\newcommand{\gev}{\text{GeV}}

\newcommand{\fb}{\text{fb}}

\newcommand{\mhpm}{ m_{H^{\pm}} }
\newcommand{\ma}{ m_{A^0} }
\newcommand{\mHH}{ m_{H^0} }
\newcommand{\mhh}{ m_{h^0} }

\newcommand{\MSQ}{ M_{\tilde{Q}} }
\newcommand{\MSU}{ M_{\tilde{U}} }
\newcommand{\MSD}{ M_{\tilde{D}} }
\newcommand{\MSL}{ M_{\tilde{L}} }
\newcommand{\MSE}{ M_{\tilde{E}} }

\newcommand{\sqrts}{\sqrt{\hat{s}}} 
\newcommand{\SM}{{\text{SM}}}
\newcommand{\MSSM}{{\text{MSSM}}}
\newcommand{\THDM}{{\text{THDM}}}
\newcommand{\Higgs}{{\text{Higgs}}}
\newcommand{\noHiggs}{{\text{no Higgs}}}
\newcommand{\SUSY}{{\text{SUSY}}}

\newcommand{\tb}{\tan\beta}

\def\NPB{{Nucl. Phys.} B}
\def\PLB{{Phys. Lett.}  B}
\def\PRL{Phys. Rev. Lett.}
 
\def\PRD{{Phys. Rev.} D}
\def\ZPC{{Z. Phys.} C}

\def\PTP{Prog.~Theor.~Phys.}

\def\Journal#1#2#3#4{{#1} {\bf #2} (#4) #3}

\begin{document}
\setlength{\unitlength}{1mm}

\begin{titlepage}
\begin{flushright}
{\bf KEK-TH-1022\\
     PITHA--05--09\\
     OU-HET 532\\
     hep-ph/0506249}
\end{flushright}
\vspace{2cm}
\begin{center}
{\Large \bf 
Enhancement of $W^\pm H^\mp$ Production at Hadron\\[.4cm]
Colliders in the Two Higgs Doublet Model}\\[2.5cm]
{\large 
{\bf Eri~Asakawa}$^a$ \footnote{E-mail: eri@yukawa.kyoto-u.ac.jp;
           Address after April 2005: YITP, Kyoto University.},
{\bf Oliver~Brein}$^b$ \footnote{E-mail: brein@physik.rwth-aachen.de}
and
{\bf Shinya~Kanemura}$^c$ \footnote{E-mail: kanemu@het.phys.sci.osaka-u.ac.jp}
}\\[0.8cm]
{\normalsize\em 
$^a$ Theory Group KEK, 
	Tsukuba, Ibaraki, 305-0801, Japan\\ 
$^b$ Institut f\"{u}r Theoretische Physik E,
	RWTH Aachen
        D-52056 Aachen, Germany\\
$^c$ Department of Physics, Osaka University, 
	Toyonaka, Osaka 560-0043, Japan}\\
\end{center}
\hfill 
\begin{abstract}  
We discuss the associated $W^\pm H^\mp$ production at 
the CERN Large Hadron Collider.
The dependence of the hadronic cross section on the Higgs sector 
parameters is investigated in detail in the framework of the 
general 
Two Higgs Doublet Model (THDM).
We study the possible enhancement of the THDM prediction
for the cross section compared to
the prediction of the Minimal Supersymmetric
Standard Model (MSSM).
We find regions in the THDM parameter space
where the THDM prediction can exceed the one of the MSSM 
by two orders of magnitude.
These regions of large cross section are in agreement with
theoretical bounds on the model, derived from the requirement
of vacuum stability and perturbative unitarity,
and are not excluded by experimental constraints.
\end{abstract}
\hfill
\end{titlepage}

\section{Introduction}

Spontaneous symmetry breakdown is a necessity in theoretical
descriptions of electroweak phenomena. We know of no other way
to unite the principle of gauge symmetry with the description
of massive vector bosons. 
While gauge symmetry is required in the theoretical 
description of electroweak physics 
in order to get meaningful predictions at the quantum level,
the existence of the massive vector bosons $W^\pm$ and $Z$
is an experimental fact since their discovery at LEP. 
Although the Standard Model (SM) of particle physics is well 
tested even at the quantum level in several cases, 
there is no sign of the Higgs boson so far. 
Up to now only bounds on the Higgs boson mass(es) could be extracted 
from high energy collision experiments.
These bounds are always obtained under the hypothesis 
that a certain model of the Higgs sector describes the data. 
However, as there is no Higgs signal yet, the choice of model
for the Higgs sector is rather unconstrained.  
Therefore, extensions of the SM Higgs sector have to be 
considered seriously in phenomenology.

Such extended Higgs sectors would also be considered as 
low energy effective theories 
of new physics models beyond the SM, which  
are proposed to describe physics at higher energies than 
the electroweak scale. 
Supersymmetry is one of the examples for such 
new physics scenarios, which requires at least two scalar
doublet fields in the Higgs sector. 
Apart from supersymmetry, there are varieties of new physics models
most of which have been proposed 
during the last decade,
such as extra dimension models \cite{ED}, 
the little Higgs models \cite{LH} and others.
The low energy effective theories for some of these 
models
predict extended Higgs sectors.
Non-minimal Higgs sectors are also considered in models 
with strongly interacting dynamics for the symmetry breaking 
such as top-color models \cite{TC}. 
Furthermore, some models designed to explain tiny neutrino masses \cite{Zee},  
CP-violation \cite{weinbergCP}, 
and electroweak baryogenesis \cite{EB} also require an
extension of the minimal Higgs sector of the SM. 

From the phenomenological point of view, such extensions of the 
Higgs sector have to meet two major restrictions from 
experiments: 
a) the electroweak rho-parameter has to 
be one up to a few per mille and 
b) large flavor changing neutral currents (FCNC) have to be absent. 
By discarding models which do not meet these criteria or meet them 
only by fine-tuned choices of model parameters, one ends up
with models which have $\rho = 1$ and no FCNC at tree-level.
In particular, it is well known that all models with 
an arbitrary number of isospin $SU(2)$-doublets 
and -singlets can be of that type \cite{HHG} by imposing 
discrete symmetries to avoid FCNC. 

The Two Higgs Doublet Model (THDM) is the model with 
the minimal extension of the SM Higgs sector, 
which leads to new phenomena
in the matter and gauge field sector\footnote{
A mere addition of scalar isospin-singlets would only modify 
the Higgs boson self-interactions.}.
In the THDM, tree level FCNC can be eliminated by imposing 
a discrete symmetry, under which there are two possibilities 
depending on its charge assignment \cite{DS,HHG}; i.e., 
(Type I) only one of the Higgs doublets gives mass to the 
fermions, and (Type II) one gives mass to the up-type quarks 
and the other to the down-type quarks and charged leptons.
The Higgs sector of the minimal supersymmetric 
extension of the SM (MSSM) is a special case of the type II THDM.  
As the general THDM may be considered as a low energy 
description of other new physics models, it 
is an interesting question whether one can distinguish 
experimentally between the 
MSSM and such a model.

The question about the new physics model underlying the Higgs
sector can be addressed only after the existence
of Higgs particles has been established experimentally.
The Higgs search program at high energy colliders 
divides basically into three steps:
a) discovery of the Higgs boson(s), 
b) measurement of fundamental properties
like its mass, width, spin, parity etc., 
and c) determination of the underlying 
model of the Higgs sector; i.e., the measurement of the Higgs bosons' 
couplings to other particles and to themselves 
and the measurement of quantum effects.
The current paper deals with an example 
which might contribute to
last step.
As is well known, the discovery of a charged Higgs boson
would be an unambiguous sign of an extended Higgs sector.
At hadron colliders, the main production mechanisms 
are top-quark decay in top-quark pair production 
if $\mhpm < m_t + m_b$ \cite{bawaetal} and single $H^-(H^+)$ 
production by bottom-gluon scattering and gluon fusion 
($g b\to t H^-$, $gg \to H^- t \bar b/H^-\bar\tau\nu_\tau$ and 
charge conjugated) if $\mhpm > m_t + m_b$ \cite{gunionetal}.

In this paper, we study $H^\pm W^\mp$ production at hadron colliders.
This 
process is not a main production process 
for the charged Higgs boson, but it turns out to be strongly model
dependent in contrast to the main production mechanisms.
After the discovery of a charged Higgs boson, the observation
of this process could potentially help to unravel the underlying 
model of the Higgs sector.
The main purpose of this paper is to demonstrate this possibility
by studying the predictions of the MSSM and THDM in comparison.
The MSSM prediction for $H^\pm W^\mp$ production at hadron colliders
has been studied by several authors 
\cite{pphw1,pphw-kniehl,BHK,zhouetal,yangetal,hollik-zhu,mo}.
Especially, the detectability of $H^\pm W^\mp$ production 
at the CERN Large Hadron Collider (LHC)
in the framework of the MSSM has been studied in Ref.~\cite{mo}
which concludes with rather poor prospects.
We reconsider this process in the THDM, 
and study discriminative features 
with respect to the MSSM.
The cross section is evaluated in the THDM in a wide 
parameter range.
We exclude areas of parameter space
by the requirement of vacuum stability \cite{VS1,VS2} and perturbative 
unitarity \cite{PU1,PU2,PU3}. \footnote{Similar approaches
are seen in Refs.\cite{simapproach}} 
Experimental results, such as 
on the rho-parameter \cite{pdg2004}, 
the muon anomalous magnetic moment \cite{amu-exp} 
and $b \to s \gamma$ \cite{bsgamma-ex} are also taken into account. 
We find that in the regions of parameter space 
which are not excluded by the theoretical requirements 
and the experimental constraints, a large enhancement 
of the hadronic cross section can be obtained 
as compared to the MSSM prediction.
Therefore, in certain cases, the signal can be detectable at the LHC.

The rest of the paper is organized as follows. 
In section 2, we briefly review the THDM and 
present the restrictions for the THDM parameters,
derived from theoretical constraints and present experimental data.
Section 3 deals with 
the essentials of the $W^\pm H^\mp$ production process
at hadron colliders. 
In Section 4, we discuss the THDM prediction for the production
cross section in comparison with the MSSM prediction,
having regard to all the parameter restrictions discussed in section 2.
Our conclusions and the appendix follow.

\section{THDM parameters}
The THDM contains two scalar weak isospin doublets $\Phi_1,\Phi_2$
with hypercharge $Y(\Phi_i)=+1$. Conventionally, there are two types 
of THDMs which are characterized by the way the scalar doublets are
coupled to fermions. In type I all fermions couple to one 
doublet, while in type II up- and down-type fermions couple to 
different doublets \cite{HHG}. In either type FCNC are automatically absent 
at tree level \cite{DS}. 
Here, we consider a THDM of type II.
The Higgs sector of the CP-conserving THDM can be described 
by the following potential \cite{HHG}:
\begin{multline}
V(\Phi_1,\Phi_2)  = \lambda_1 ( \Phi_1^\dagger \Phi_1 -v_1^2)^2
        +\lambda_2 (\Phi_2^\dagger \Phi_2-v_2^2)^2
+\lambda_3 \big[ (\Phi_1^\dagger \Phi_1-v_1^2)
        +(\Phi_2^\dagger \Phi_2-v_2^2) \big]^2\\
+\lambda_4 \big[ (\Phi_1^\dagger \Phi_1)(\Phi_2^\dagger \Phi_2)
        - (\Phi_1^\dagger \Phi_2)(\Phi_2^\dagger \Phi_1) \big]
+\lambda_5 \big[ \text{Re}(\Phi_1^\dagger \Phi_2) - v_1 v_2\big]^2
+\lambda_6 \big[ \text{Im}(\Phi_1^\dagger \Phi_2)\big]^2 \;,
\label{2hdm-higgs-pot}
\end{multline}
with two real parameters with mass dimension one ($v_1,v_2$) 
and 6 dimensionless real parameters ($\lambda_1,\cdots,\lambda_6$).
This potential has its minimum at $\Phi_i=(0, v_i)^T (i=1,2)$.
Re-expressing the fields such that the new degrees of freedom vanish
at the minimum and diagonalizing the resulting bilinear scalar
interaction terms one obtains a separation of the physical and 
unphysical spectrum of the model. 
The physical spectrum of the Higgs sector 
of the CP-conserving THDM consists of 3 neutral 
Higgs bosons, 2 CP-even ($h^0,H^0$) and one CP-odd ($A^0$),
and two charged ones ($H^+,H^-$). The 8 free parameters of
the Higgs sector 
can also be chosen to be 
the modulus of the vacuum expectation value 
$v = \sqrt{v_1^2+v_2^2} = \sqrt{2^{-3/2} G_F^{-1}}$,
the masses of the Higgs bosons,
$m_{h^0}, m_{H^0}, m_{A^0}$ and $\mhpm$, 
the mixing angles $\alpha$ and $\beta$ 
used in the 
diagonalization of the Higgs boson propagator matrices,
and 
$\lambda_5$.
The discrete symmetry of the Higgs potential 
(e.g. $\Phi_1\to\Phi_1$, $\Phi_2\to -\Phi_2$) 
is broken by a mass-dimension two term 
proportional to 
$M^2=v^2\lambda_5$.
The physical meaning of $M$ is the cut-off scale of the effective SM
when $M\gg v$.
In the following, we review the constraints which we take into account in our 
numerical study.\smallskip

\noindent {\bf 1. theoretical constraints}\\
The coupling constants in Eq.~(\ref{2hdm-higgs-pot}) can be
restricted by imposing theoretical requirements for  
the consistency of the model.
Here, we use the conditions derived from the requirement of 
vacuum stability \cite{VS1,VS2}
and 
perturbative unitarity
\cite{PU1,PU2,PU3}
for the tree-level coupling constants. 
The condition of vacuum stability is given by 
\begin{align}
\nonumber
& \lambda_1 + \lambda_3 > 0\;,\quad\quad \lambda_2 + \lambda_3 > 0\;, \\
& 2\sqrt{(\lambda_1 + \lambda_3)(\lambda_2 + \lambda_3)} 
+2\lambda_{3} +\lambda_4
+ {\text{min}}
\left[ 0, \lambda_{5} - \lambda_{4} , \lambda_{6}-\lambda_{4}
 \right] >0\;.  
\label{eq:VS}
\end{align} 
The requirement of perturbative unitarity 
demands that the magnitudes of all tree-level S-wave amplitudes for the
elastic scattering of longitudinally polarized
gauge and Higgs bosons stay within the limit
set by unitarity. 
In our analysis, we consider the 14 neutral channels \cite{PU2}.
The expressions for the eigenvalues of the scattering matrix, 
$a_i$ $(i=1,\ldots,14)$, are summarized in Appendix \ref{s-wave-amp}.
\smallskip

\noindent {\bf 2. rho-parameter constraint}\\
The electroweak rho-parameter, which is one at tree-level in the SM,
is a scheme-dependent quantity beyond the leading order in
perturbation theory .
For models beyond the SM the definition 
$\rho = \rho_0 \rho_\SM = m_W^2/(m_Z^2  c_w^2)$
is used in Ref. \cite{pdg2004} in the $\overline{MS}$ scheme, 
where $\rho_\SM$ absorbs all SM radiative corrections 
and $\rho_0$ parameterizes the extra new physics contributions,
i.e. $\rho_0=1$ if the new physics contributions vanish.
A value for $\rho_0$ has been obtained
from a global fit to electroweak precision observables \cite{pdg2004},
\begin{align}\label{rho0-fit}
\rho_0 & = 0.9998^{+0.0025}_{-0.0010} \;,
\end{align} 
where the error bar given above is at the $2 \sigma$ level.
The only difference between the SM and the THDM is the Higgs sector.
Thus, we can decompose the contributions to $\rho (\equiv \rho_\THDM)$
into all contributions containing at least one virtual Higgs boson,
$\rho_{\THDM,\Higgs}$, and all others, which coincide with the SM
contributions:
$
\rho_\THDM  = \rho_{\SM,\noHiggs} + \rho_{\THDM,\Higgs}
$.
Likewise, the SM rho-parameter decomposes into 
$
\rho_\SM = \rho_{\SM,\noHiggs} + \rho_{\SM,\Higgs} 
$.
Hence, the deviation $\delta\rho_0 = \rho_0 - 1$ at the one-loop level
is just the difference
$
\delta\rho_0 = \rho_{\THDM,\Higgs} - \rho_{\SM,\Higgs}
$.
In our study, we constrain the THDM parameters such that the one-loop prediction
for $\delta\rho_0$ stays within the range indicated by Eq. (\ref{rho0-fit}); 
i.e.,  $-0.0012\leq\delta\rho_0\leq 0.0023$.
The specific formulas for this difference can be found 
in Refs. \cite{HHG,rho0-formulas}.
\smallskip

\noindent {\bf 3. constraint from $a_\mu$}\\
The latest results on the measurement of the 
anomalous magnetic moment of the muon \cite{amu-exp},
$a_\mu$, suggest that the difference between measurement 
and SM prediction is \cite{pdg2004}:
\begin{align}\label{Delta-amu}
\Delta a_\mu & :=  a_\mu^{\text{exp.}} - a_\mu^{\text{SM}} 
= (25.7 \pm 8.54 \pm 4.90)\E{-10}
\;,
\end{align}
where the first error is the total experimental error 
and the second is the theoretical uncertainty of the SM prediction.
In the THDM, radiative corrections to processes 
without any external Higgs bosons
always split into contributions
without any virtual Higgs boson 
and others which involve at least one virtual Higgs boson.
Thus, the difference between the THDM and the SM prediction for 
$a_\mu$ is given by the difference in the Higgs sector contribution,
$
\delta a_\mu  =  a_\mu^{\THDM,\Higgs} - a_\mu^{\SM,\Higgs} 
$.
As is well known, the leading order virtual Higgs contributions 
are suppressed by two Higgs-muon Yukawa couplings \cite{one-loop-damu}.
At the two-loop level, a virtual Higgs boson can couple to the muon
and one internal loop of heavy particles, e.g. Fermions with a 
much larger Yukawa coupling.
This is the class of so-called ``Barr-Zee'' type Feynman graphs \cite{barr-zee-type}.
Their contributions to $a_\mu$ exceed the leading order virtual Higgs contributions
by several orders of magnitude.
In our calculation of $\delta a_\mu$, 
we include all relevant one-loop contributions \cite{one-loop-damu}
and all Barr-Zee type two-loop contributions with a closed fermion 
or charged Higgs boson loop
\cite{two-loop-damu}. 
Assuming the validity of the THDM, we constrain its parameter space such that
the value of 
$\delta a_\mu$ stays close to $\Delta a_\mu$ within the $2\sigma$ error bars;
i.e., $-1.2\E{-10}\leq\delta a_\mu \leq 52.6\E{-10}$.
\smallskip

\noindent {\bf 4. constraint from $b \to s \gamma$}\\
In our discussion, we take a charged Higgs boson with $\mhpm = 400\,\gev$.
Therefore, the limits for new physics contributions to the decay $b \to s \gamma$
are respected \cite{bsgamma-ex,bsgamma} 
for all values of $\tb$ which we discuss.

\section{$W^\pm H^\mp$ production at hadron colliders}

\noindent {\bf Partonic processes}\\
In the framework of the parton model, there are 
two distinct subprocesses
which contribute to the production of 
a charged Higgs boson, $H^+$, and a electroweak gauge boson, $W^-$:
$b \bar b$-annihilation,
\begin{align}\label{bbhw}
b(k,\alpha,\sigma) + \bar b(\bar k, \beta, \bar\sigma) \to W^-(p,\lambda) + H^+(\bar p)
\;,
\end{align}
and gluon fusion,
\begin{align}\label{gghw}
g(k,a,\sigma) + g(\bar k, b, \bar\sigma) \to W^-(p,\lambda) + H^+(\bar p)
\;.
\end{align}
In- and outgoing momenta of the initial and final state particles are 
denoted by $k, \bar k$ and $p, \bar p$ respectively, the helicities of the 
initial state gluons or quarks by $\sigma, \bar\sigma$ and of the final
state $W$ by $\lambda$. $a,b$ denote the gluon SU(3)-color 
indices 
and $\alpha, \beta$ the quark color indices.
The square of the center of mass energy of the parton system is then given by
$\hat s = (k+\bar k)^2  = (p + \bar p)^2$.

\begin{figure}[t]
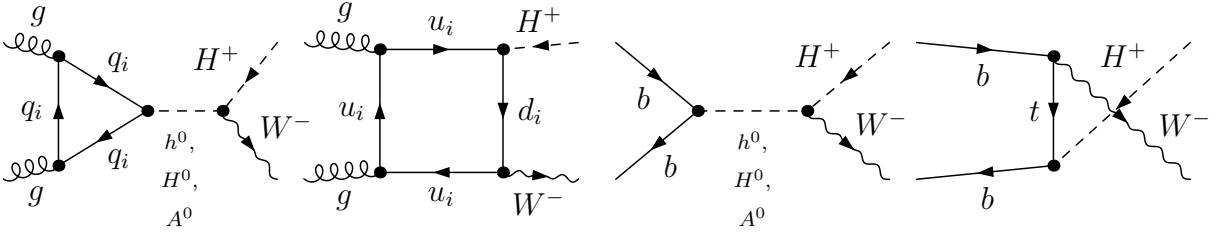

\begin{feynartspicture}(80,40)(2,1)
\FADiagram{}
\FAProp(0.,15.)(4.,14.)(0.,){/Cycles}{0}
\FALabel(2.54571,16.2028)[b]{$g$}
\FAProp(0.,5.)(4.,6.)(0.,){/Cycles}{0}
\FALabel(2.37593,4.47628)[t]{$g$}
\FAProp(20.,15.)(16.,10.)(0.,){/ScalarDash}{1}
\FALabel(17.2697,12.9883)[br]{$H^+$}
\FAProp(20.,5.)(16.,10.)(0.,){/Sine}{-1}
\FALabel(18.7303,7.98828)[bl]{$W^-$}
\FAProp(16.,10.)(10.5,10.)(0.,){/ScalarDash}{0}
\FALabel(13.25,9.18)[t]{$\begin{array}[1]{c}
	\scriptstyle h^0,\\
	\scriptstyle H^0,\\
	\scriptstyle A^0
			\end{array}$
			}
\FAProp(4.,14.)(4.,6.)(0.,){/Straight}{-1}
\FALabel(2.93,10.)[r]{$q_i$}
\FAProp(4.,14.)(10.5,10.)(0.,){/Straight}{1}
\FALabel(7.58235,12.8401)[bl]{$q_i$}
\FAProp(4.,6.)(10.5,10.)(0.,){/Straight}{-1}
\FALabel(7.58235,7.15993)[tl]{$q_i$}
\FAVert(4.,14.){0}
\FAVert(4.,6.){0}
\FAVert(16.,10.){0}
\FAVert(10.5,10.){0}

\FADiagram{}
\FAProp(0.,15.)(5.5,14.5)(0.,){/Cycles}{0}
\FALabel(2.95371,16.5108)[b]{$g$}
\FAProp(0.,5.)(5.5,5.5)(0.,){/Cycles}{0}
\FALabel(2.89033,4.18637)[t]{$g$}
\FAProp(20.,15.)(14.5,14.5)(0.,){/ScalarDash}{1}
\FALabel(17.1097,15.8136)[b]{$H^+$}
\FAProp(20.,5.)(14.5,5.5)(0.,){/Sine}{-1}
\FALabel(17.1097,4.18637)[t]{$W^-$}
\FAProp(5.5,14.5)(5.5,5.5)(0.,){/Straight}{-1}
\FALabel(4.43,10.)[r]{$u_i$}
\FAProp(5.5,14.5)(14.5,14.5)(0.,){/Straight}{1}
\FALabel(10.,15.57)[b]{$u_i$}
\FAProp(5.5,5.5)(14.5,5.5)(0.,){/Straight}{-1}
\FALabel(10.,4.43)[t]{$u_i$}
\FAProp(14.5,14.5)(14.5,5.5)(0.,){/Straight}{1}
\FALabel(15.57,10.)[l]{$d_i$}
\FAVert(5.5,14.5){0}
\FAVert(5.5,5.5){0}
\FAVert(14.5,14.5){0}
\FAVert(14.5,5.5){0}
\end{feynartspicture}
\begin{feynartspicture}(80,40)(2,1)
\FADiagram{}
\FAProp(0.,15.)(6.,10.)(0.,){/Straight}{1}
\FALabel(2.48771,11.7893)[tr]{$b$}
\FAProp(0.,5.)(6.,10.)(0.,){/Straight}{-1}
\FALabel(3.51229,6.78926)[tl]{$b$}
\FAProp(20.,15.)(14.,10.)(0.,){/ScalarDash}{1}
\FALabel(16.4877,13.2107)[br]{$H^+$}
\FAProp(20.,5.)(14.,10.)(0.,){/Sine}{-1}
\FALabel(17.5123,8.21074)[bl]{$W^-$}
\FAProp(6.,10.)(14.,10.)(0.,){/ScalarDash}{0}
\FALabel(10.,9.18)[t]{$\begin{array}[1]{c}
	\scriptstyle h^0,\\
	\scriptstyle H^0,\\
	\scriptstyle A^0
			\end{array}$}
\FAVert(6.,10.){0}
\FAVert(14.,10.){0}

\FADiagram{}
\FAProp(0.,15.)(10.,14.)(0.,){/Straight}{1}
\FALabel(4.84577,13.4377)[t]{$b$}
\FAProp(0.,5.)(10.,6.)(0.,){/Straight}{-1}
\FALabel(5.15423,4.43769)[t]{$b$}
\FAProp(20.,15.)(10.,6.)(0.,){/ScalarDash}{1}
\FALabel(16.8128,13.2058)[br]{$H^+$}
\FAProp(20.,5.)(10.,14.)(0.,){/Sine}{-1}
\FALabel(17.6872,8.20582)[bl]{$W^-$}
\FAProp(10.,14.)(10.,6.)(0.,){/Straight}{1}
\FALabel(9.03,10.)[r]{$t$}
\FAVert(10.,14.){0}
\FAVert(10.,6.){0}
\end{feynartspicture}
\caption{\label{hw-graphs}
Typical Feynman graphs for the partonic 
processes gluon fusion and $b \bar b$ annihilation contributing
to $W^\pm H^\mp$ production at a hadron collider. For gluon fusion
there are in total two triangle-type topologies for each quark flavor
and six box-type topologies for each quark generation.
}
\end{figure}

The leading order $b \bar b$-annihilation amplitude (\ref{bbhw})
consists of two types of tree-level Feynman graphs:
(i) graphs with s-channel exchange of a neutral Higgs boson
and
(ii) graphs with a t-channel virtual top quark (see Fig. \ref{hw-graphs}).
The amplitude for the gluon-fusion process (\ref{gghw}) 
is given in leading order in perturbation theory 
by a set of one-loop Feynman graphs (see Fig.\ref{hw-graphs}).
Gluon fusion, though loop-induced, may contribute 
significantly to the cross section, because of the large number 
of gluon-gluon collisions with sufficient center-of-mass energy
to exceed the production threshold 
at high energy hadron colliders.

Associated $W^\pm H^\mp$ production
via $b\overline{b}$ annihilation and
the gluon fusion have been discussed at first 
for a THDM with MSSM parameter values including the loop contributions
from top and bottom quarks in the approximation $m_b = 0$ \cite{pphw1}. 
Therefore, this study only covered MSSM scenarios with small $\tan\beta$ and 
scalar quark masses which
are sufficiently heavy to decouple from the loop contributions.
This work has been extended 
by
including squark-loop contributions and 
a non-zero $b$-quark mass
\cite{pphw-kniehl,BHK}, 
thus allowing the investigation  
of the process for arbitrary scalar quark masses and 
values of $\tan\beta$.
The gluon-fusion channel in the MSSM has also been studied in Ref. \cite{zhouetal}
and for the $b\bar b$-annihilation channel 
the supersymmetric electroweak corrections \cite{yangetal}
and the QCD corrections \cite{hollik-zhu} at one-loop order are known.
A phenomenological study of the signal-to-background ratio for the 
semi-leptonic signature
$W^\pm H^\mp \to W^\pm t b \to b\bar b W^+W^- \to b\bar b j j l 
+\text{missing energy/momentum}$ for a MSSM-like THDM (neglecting the
box-loop contributions) has been performed in Ref. \cite{mo}.
The authors of Ref. \cite{mo} showed that the background rate, 
which mainly comes from $t\overline{t}$ production,
overwhelms the signal rate by two to three orders of magnitude.
A generic feature of the MSSM gluon-fusion amplitude is a strong
negative interference between triangle- and box-type quark-loop
Feynman graphs.
This behavior, already noted in Ref.~\cite{pphw1}, 
is due to the relations between Higgs masses in the MSSM, 
and leads to a much smaller cross section than one would 
obtain by squaring triangle- or box-type quark-loops alone.

The
$W^\pm H^\mp$ production cross section at hadron colliders 
in the framework of the general THDM
has not been studied in the literature\footnote{
A preliminary study of the THDM gluon-fusion process
has been presented in Ref. \cite{DRA}.
}.
We discuss this cross section in the framework of a general
CP-conserving type II THDM, which contains the MSSM Higgs
sector as a special case and therfore allows for straightforward
comparisons between the models.
Dropping the superpartner contributions and allowing the
Higgs boson masses ($m_{h^0}$, $m_{H^0}$, $m_{A^0}$, $\mhpm$) 
and the mixing angle in the Higgs sector ($\alpha$)
to be free parameters, one obtains
the expression for the partonic cross section 
$\hat\sigma_{gg\to W^- H^+ }$ in the THDM 
from the results of Ref.~\cite{BHK}. 
The formulas for $\hat\sigma_{b\bar b\to W^- H^+ }$ in the THDM
can be taken over e.g. from the MSSM calculation in Ref. \cite{pphw-kniehl}.
As all Higgs boson masses in the THDM are free parameters,
it may occur that their values lie above the threshold 
$m_W+\mhpm$. This gives rise to resonant s-channel contributions.
Therefore, we calculate the widths $\Gamma_\Phi$ ($\Phi=h^0, H^0, A^0$) 
of the neutral Higgs bosons
in leading order approximation and introduce them
in the s-channel propagators
of the Feynman graphs of the $b\bar b$-annihilation and gluon-fusion process .
Specifically, we make the following replacements for the 
propagator terms of 
neutral Higgs bosons 
in the formulas of Refs. \cite{pphw-kniehl,BHK}:
\begin{align*}
\frac{1}{\hat s - m_\Phi^2} & \to \frac{1}{\hat s - m_\Phi^2 + i m_\Phi \Gamma_\Phi}
\;.
\end{align*}
In the THDM, 
there are parameter combinations in which one Higgs boson
can decay into two others which gives a large contribution to its
width.
We take this into account by including
all self-interaction contributions to the total width of the 
neutral Higgs bosons.
The partial widths of neutral Higgs bosons
associated with decays into SM particles 
are well known and have been taken over from Ref. \cite{HHG} and 
the ones associated with
decays into Higgs particles are given by
\begin{align*}
\Gamma(\Phi_1\to\Phi_i\Phi_j) & = 
\frac{\sqrt{m_1^4+m_i^4+m_j^4-2m_1^2m_i^2-2m_1^2m_j^2-2m_i^2m_j^2}}{
16\pi\,m_1^3(1+\delta_{ij}) }
|g_{\Phi_1 \Phi_i \Phi_j}|^2,
\end{align*}
with the Higgs self-couplings $g_{\Phi_1 \Phi_i \Phi_j}$ 
listed in appendix \ref{tripleHcouplings}.
\smallskip

\noindent We describe above how to obtain the expressions for the partonic 
cross sections from the formulas of Refs. \cite{pphw-kniehl,BHK}.
However, our calculation of the partonic cross sections 
has been performed independently with the help of 
the computer programs Feyn\-Arts and Form\-Calc \cite{FAFC}.

\noindent {\bf Hadronic cross section}\\
The hadronic inclusive cross section for $W^- H^+$ production in 
proton-proton collisions at a total hadronic center of mass energy 
$\sqrt{S}$ can be written as a convolution \cite{pQCD-lect},
\begin{align}
\label{hadronx}
\sigma_{p p  \to W^- H^+ + X} & =
\sum_{\{n,m\}}^{} 
\int_{\tau_0}^{1} d\tau \frac{ d{\cal L}^{pp}_{nm} }{ d\tau }
\;\hat\sigma_{n m \to W^- H^+}(\tau S,\alpha_S(\mu_R))
= \sum_{\{n,m\}}^{} \int_{\sqrt{\hat s_0}}^{\sqrt{S}} d\sqrts
\frac{d \sigma_{nm}}{d \sqrts}
\;,
\end{align}
with the parton luminosity
\begin{align}
\frac{ d{\cal L}^{pp}_{nm} }{ d\tau } & =
        \int_{\tau}^{1} \frac{dx}{x}
        \frac{1}{1+\delta_{nm}} 
        \Big[
        f_{n/p} (x,\mu_F) f_{m/p} (\frac{\tau}{x},\mu_F)
        + f_{m/p} (x,\mu_F) f_{n/p} (\frac{\tau}{x},\mu_F)
        \Big]
        \, ,
\end{align}
where $f_{n/p} (x,\mu_F)$ denotes the density of partons of type $n$
in the proton carrying a fraction $x$ of the proton momentum
at the scale $\mu_F$. 
In our case,
there are two parton subprocesses
contributing to inclusive $W^- H^+$ hadroproduction; gluon fusion
and $b \bar{b}$-annihilation. 
The numerical evaluation has been carried out with the leading order MRST
parton distribution functions \cite{MRST} and 
with the renormalization and factorization scale $\mu_R$ and $\mu_F$ 
chosen equal to the threshold, $m_W+\mhpm$.

\section{Numerical Results}

We take $m_Z^{}$, $m_W^{}$ and $G_F^{}$ 
as the input electroweak parameters, and use values 
$m_Z=91.1876$ GeV, $m_W=80.424$ GeV and $G_F=1.16637 \times 10^{-5}$ 
GeV$^{-2}$~\cite{pdg2004}.
For the strong coupling constant $\alpha_S(\mu_R)$, we use the formula 
including the two-loop QCD corrections for $n_f=5$ with 
$\Lambda^5_{QCD}=174$ MeV which can be found in~\cite{pdg2004}.
The mass of the top and bottom quarks are fixed here as $m_t=174.3$ GeV 
and $m_b=4.7$ GeV. 

In our exemplary 
discussion we take one MSSM scenario as a reference, 
characterized by the following settings of MSSM parameters:
\begin{align}
\nonumber
M_\SUSY & = \MSQ = \MSU = \MSD = \MSL = \MSE = 1000\,\gev\;,\\
\label{ref-scenario}
\mu & = 300\,\gev\;,\\
\nonumber
X_t & = A_t - \mu \cot\beta = -1000\,\gev\;,\\
\nonumber
A_u & = A_d = A_c = A_s = A_b = 0\;,
\end{align}
where $M_\SUSY$ is a common squark mass scale, $\mu$ is the 
Higgs-superfield mass parameter in the superpotential,
$X_t$ the mixing parameter in the stop sector,
and the $A_q$ are trilinear Higgs couplings to squarks
\footnote{For more details on these
parameters in our convention see Ref. \cite{BHK}.}.
We set $\mhpm = 400\,\gev$ throughout the paper
and calculate the mass 
of the MSSM CP-odd Higgs boson using the tree-level relation $\ma^2 = \mhpm^2 - m_W^2$.
A THDM with all parameters set to the MSSM values, especially $M=m_{A^0}$,
is called MSSM-like.
In Table \ref{table}, we list the MSSM values 
for the Higgs masses and the mixing angle $\alpha$. Note that the case $\tb=1.5$
in the MSSM is actually ruled out by the latest LEP direct search results
\cite{LEP-Higgs}.
\begin{table}[t]
\begin{center}
\begin{tabular}[5]{lrrrr}
\hline\hline
& \multicolumn{4}{c}{$\tan\beta$}\\
\hline
	& 1.5     & 3       & 6     & 10\\
\hline
$\mhh\,[\gev]$ & $85.6$ & $105.5$ & $115.2$ & $117.6$ \\
\hline
$\mHH\,[\gev]$ & $404.7$ & $396.8$ & $393.3$ & $392.4$ \\
\hline
$\ma\,[\gev]$  & $391.8$ & $391.8$ & $391.8$ & $391.8$ \\
\hline
$\mhpm\,[\gev]$ & $400.0$ & $400.0$ & $400.0$ & $400.0$ \\
\hline
$\alpha\, [\text{rad}]$ & $-0.63465$ & $-0.36335$ & $-0.19050$ & $-0.11558$\\
\hline
$\sigma_{pp\to W^\pm H^\mp}^\MSSM [\fb]$ & $37.78$ & $9.48$ & $3.08$ & $2.96$\\
\hline\hline
\end{tabular}
\caption{\label{table} MSSM values for the Higgs masses and the mixing angle $\alpha$.
	The prediction for the hadronic cross section in our MSSM reference 
	scenario (see Eq. (\ref{ref-scenario})) are also shown.
        }
\end{center}
\end{table}

\begin{figure}[t]
{\setlength{\unitlength}{1cm}
\begin{picture}(14,6)(0,0)
\psfrag{SQRTS}[l][l]{$\scriptstyle \sqrts\;[\gev]$}
\psfrag{SIGMA}[l][l]{$\scriptstyle \sigma\;[\fb]$}
\put(-.5,0){\resizebox{1.05\width}{1.05\height}{\includegraphics*{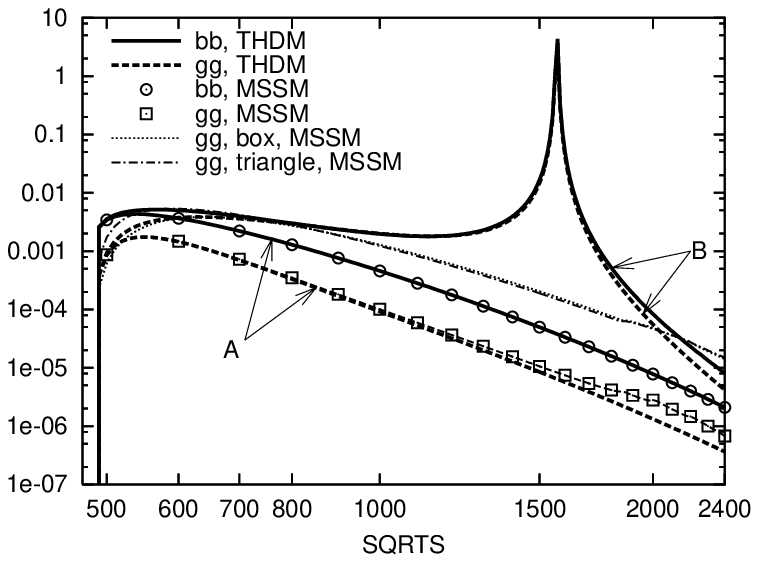}}}
\put(7.6,0){\resizebox{1.05\width}{1.05\height}{\includegraphics*{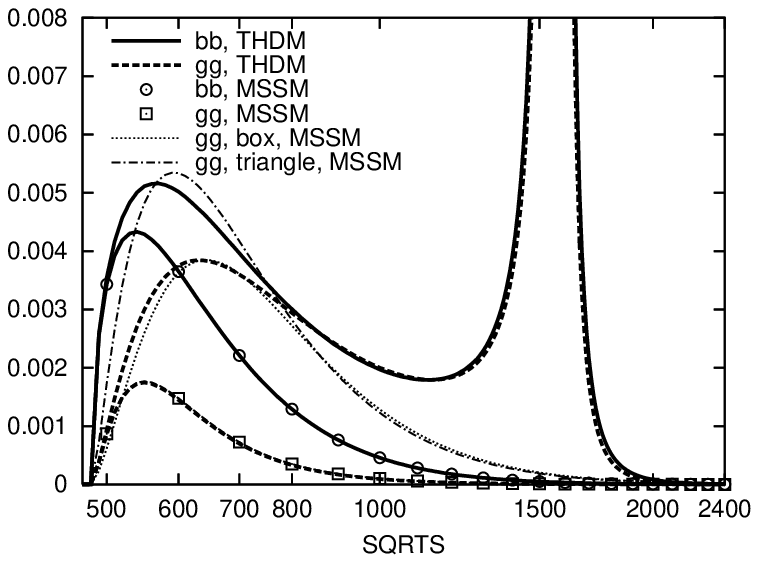}}}
\end{picture}
}
\caption{\label{diffsigmahad-fig} The differential hadronic cross section 
$d\sigma_{nm}/d\sqrts$ in $\fb/\gev$ 
for $W^- H^+$ production
is plotted versus $\sqrts$.
Thick and thin lines show THDM and MSSM results respectively.
Dashed and solid lines show the $gg$ and $b \bar b$ 
contribution respectively.
Two THDM scenarios are displayed: one with all parameters MSSM-like (A)
and one with $m_A = 4 m_A^{\text{MSSM}}$ (B).
The MSSM prediction is highlighted by circles ($b\bar b$) 
and squares ($gg$).
Also shown is the MSSM ``prediction'' for the 
$gg$ cross section only using box- (dotted lines) or
triangle-type (dot-dashed) Feynman graphs in the calculation.
}
\end{figure}

Fig. \ref{diffsigmahad-fig} shows a comparison between the THDM and 
MSSM predictions for the differential hadronic cross section
\begin{align}
\frac{d \sigma_{n m}}{d \sqrts} & = \frac{2\sqrts}{S}
\left. \frac{d {\cal L}_{nm}^{pp}}{d \tau} 
\right|_{\tau=\frac{\sqrts}{S}}
\hat\sigma_{nm\to W^- H^+}(\sqrts, \alpha_S(\mu_R))
\;,
\end{align}
as a function of 
$\sqrts$.
In the linear plot on the right hand side, one 
can see that the $gg$- and $b \bar b$-channel contribution
to the hadronic cross section is of comparable 
size, whereas the partonic cross section 
$\hat\sigma_{gg\to W^- H^+}$
is orders of magnitude lower than 
$\hat\sigma_{b\bar b\to W^- H^+}$.
This is caused by the enhancement factor
due to the large number of gluon-gluon collisions at high
energy hadron colliders. 
In Fig. \ref{diffsigmahad-fig}, two THDM scenarios are displayed:
one (scenario A) is completely MSSM-like, with the settings
$\mhpm = 400\,\gev$ and $\tb = 6$, 
and the the other (scenario B) coincides with the first one except for the
choice $m_{A^0} = 4 m_{A^0}^{\text{MSSM}}$, which leads to the 
peak at $\sqrts = m_{A^0} \approx 1600\,\gev$.
We also show in Fig. \ref{diffsigmahad-fig} the MSSM prediction by thin solid and 
dashed lines further marked by circles and boxes respectively.
Clearly, there is almost no deviation from the MSSM-like
THDM except for squark threshold effects around 
$\sqrts \approx 2000\,\gev$ which can be seen in the left plot.

In order to demonstrate the strong negative interference 
in the MSSM gluon-fusion process, we display artificial
MSSM predictions for this process using only
Feynman graphs 
either of
box-type (thin dotted lines) or of triangle-type (thin dot-dashed lines)
in the calculation. Those are more than one 
order of magnitude larger than the full result over a wide 
range of $\sqrts$ (see Fig. \ref{diffsigmahad-fig}, left plot).
From Fig. \ref{diffsigmahad-fig}, we learn that we can get a much larger 
cross section near to the production-threshold region by
near-decoupling of the triangle-type  graphs. In this region
the cross section in the THDM scenario B 
is close to the ``only box graphs'' MSSM result.
However, the resonant peak far off the production threshold also
contributes significantly to the total hadronic cross section.
Naively, one could have thought the peak far off
would not contribute significantly because of the much lower 
parton luminosity.

\begin{figure}[ht]
{\setlength{\unitlength}{1cm}
\begin{picture}(15.5,15.8)(-.7,0)
\psfrag{SQRTS}[l][l]{$\scriptstyle \sqrts\;[\gev]$}
\psfrag{SIGMA}[l][l]{$\scriptstyle \sigma\;[\fb]$}
\put(.1,6.5){$\tb = 6$}
\put(2.5,0.1){$m_{A^0} [\gev]$}
\put(5.2,7.4){$m_{H^0} [\gev]$}
\put(-5.5,-2){\resizebox{1.1\width}{1.1\height}{
\includegraphics*{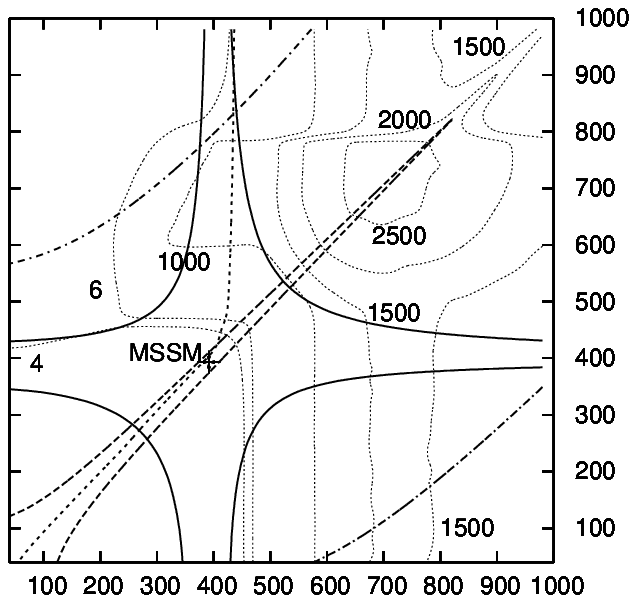}}}
\put(8.1,6.5){$\tb = 10$}
\put(10.5,0.1){$m_{A^0} [\gev]$}
\put(13.2,7.4){$m_{H^0} [\gev]$}
\put(2.5,-2){\resizebox{1.1\width}{1.1\height}{
\includegraphics*{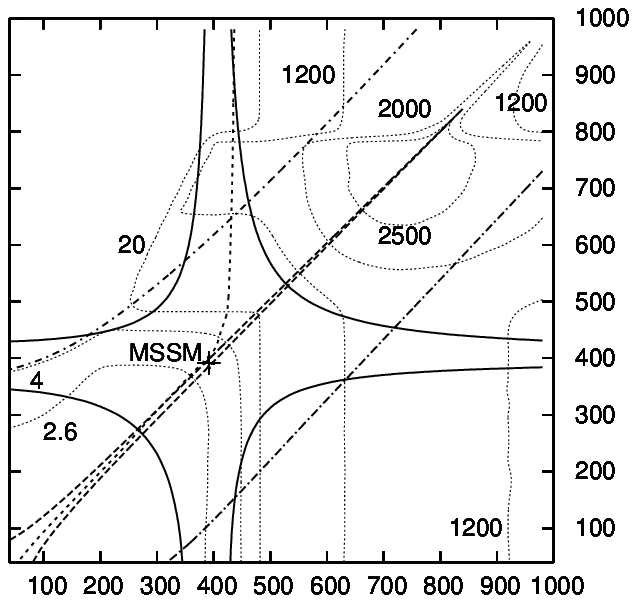}}}
\put(0.1,14.5){$\tb = 1.5$}
\put(10.5,8.1){$m_{A^0} [\gev]$}
\put(13.2,15.4){$m_{H^0} [\gev]$}
\put(-5.5,6){\resizebox{1.1\width}{1.1\height}{
\includegraphics*{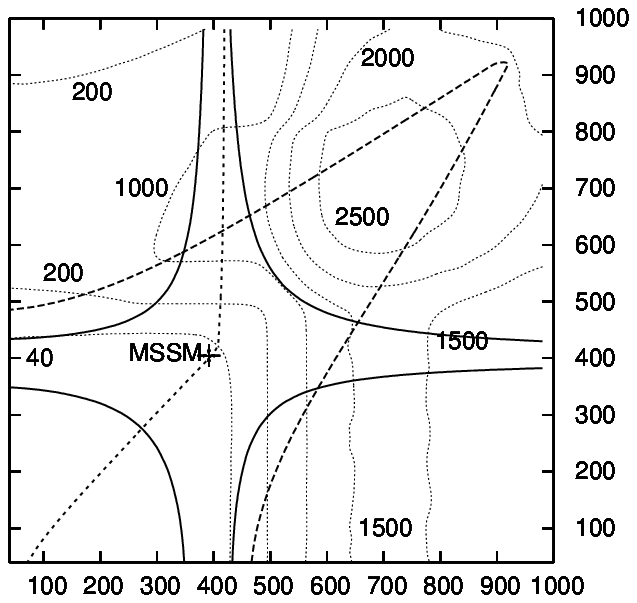}}}
\put(8.1,14.5){$\tb = 3$}
\put(2.5,8.1){$m_{A^0} [\gev]$}
\put(5.2,15.4){$m_{H^0} [\gev]$}
\put(2.5,6){\resizebox{1.1\width}{1.1\height}{
\includegraphics*{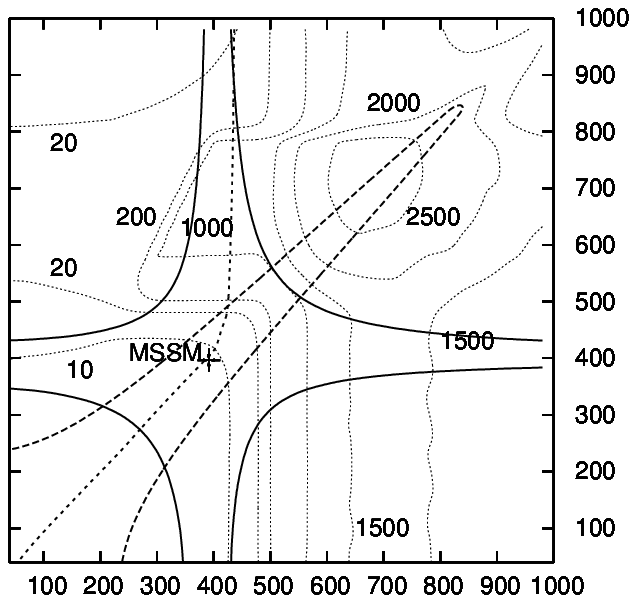}}}
\end{picture}
}
\caption{\label{m2-ma2-case}
Contour of the Hadronic cross section $\sigma_{pp\to W^\pm H^\mp}$ 
in fb for the case $M^2=m_{A^0}^2$
(thin dotted lines). 
The limiting contours 
for $\delta\rho_0$ (thick solid lines) allow the cross-shaped area, 
and 
for $\delta a_\mu$ (thick dot-dashed lines) the whole displayed 
area for $\tb=1.5$, $3$ and the band-shaped area for $\tb=6$, $10$.
The perturbative unitarity constraint, 
$\text{max}\,|a_i| < 1/2$, (thick dashed lines) allows 
the pointed area which includes the lower left corner.
The area allowed by 
the vacuum stability condition (\ref{eq:VS}) 
is left of the thick short-dashed contour.
The cross, labeled ``MSSM'', shows the point where 
all Higgs sector parameters coincide with the ones
of our reference MSSM scenario.
}
\end{figure}

\begin{figure}[ht]
{\setlength{\unitlength}{1cm}
\begin{picture}(15.5,15.8)(-.7,0)
\psfrag{SQRTS}[l][l]{$\scriptstyle \sqrts\;[\gev]$}
\psfrag{SIGMA}[l][l]{$\scriptstyle \sigma\;[\fb]$}
\put(.1,6.5){$\tb = 6$}
\put(2.5,0.1){$m_{A^0} [\gev]$}
\put(5.2,7.4){$m_{H^0} [\gev]$}
\put(-5.5,-2){\resizebox{1.1\width}{1.1\height}{
\includegraphics*{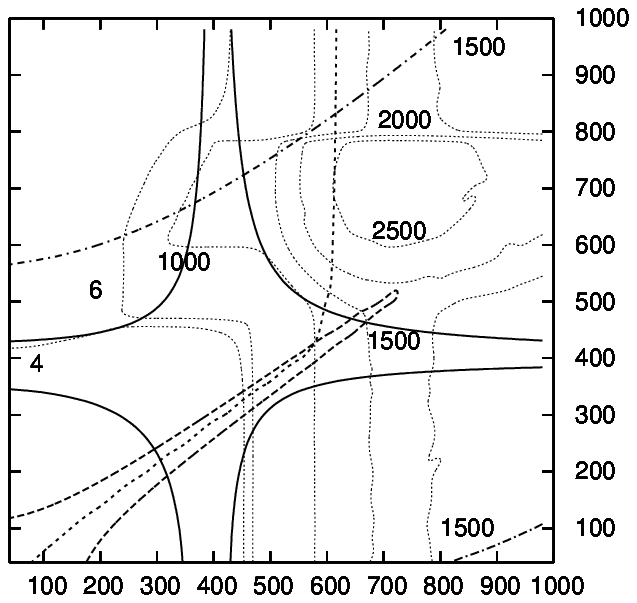}}}
\put(8.1,6.5){$\tb = 10$}
\put(10.5,0.1){$m_{A^0} [\gev]$}
\put(13.2,7.4){$m_{H^0} [\gev]$}
\put(2.5,-2){\resizebox{1.1\width}{1.1\height}{
\includegraphics*{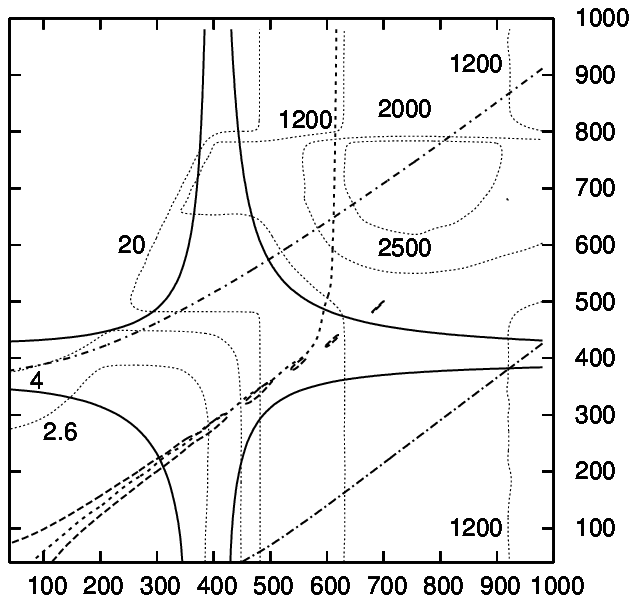}}}
\put(0.1,14.5){$\tb = 1.5$}
\put(10.5,8.1){$m_{A^0} [\gev]$}
\put(13.2,15.4){$m_{H^0} [\gev]$}
\put(-5.5,6){\resizebox{1.1\width}{1.1\height}{
\includegraphics*{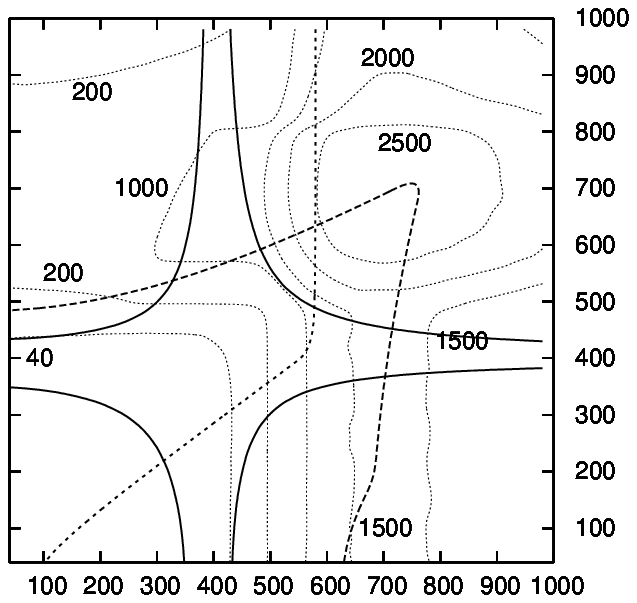}}}
\put(8.1,14.5){$\tb = 3$}
\put(2.5,8.1){$m_{A^0} [\gev]$}
\put(5.2,15.4){$m_{H^0} [\gev]$}
\put(2.5,6){\resizebox{1.1\width}{1.1\height}{
\includegraphics*{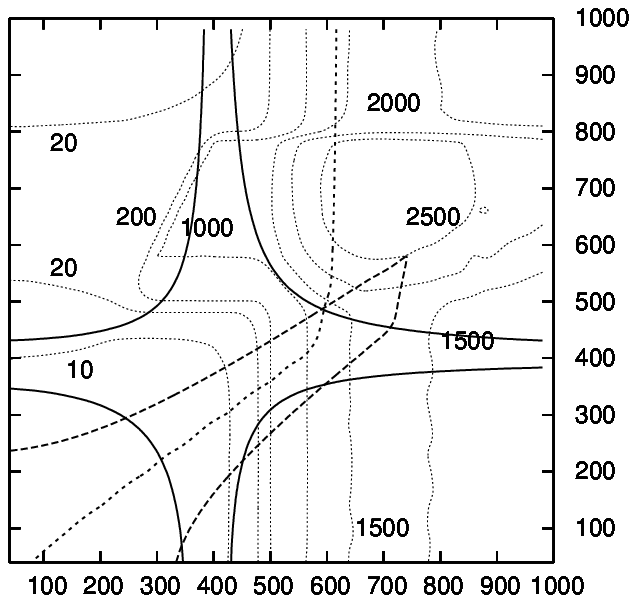}}}
\end{picture}
}
\caption{\label{m2-halfma2-case}
Contour of the Hadronic cross section $\sigma_{pp\to W^\pm H^\mp}$ 
in fb for the case $M^2=m_{A^0}^2/2$
(thin dotted lines). 
The area of allowed parameter space, defined by the superposition
of limiting contours for $\delta\rho_0$ (thick solid lines),
$\delta a_\mu$ (thick dot-dashed lines),
perturbative unitarity (thick dashed lines),
and vacuum stability (thick short-dashed lines), 
is described in Fig.~\ref{m2-ma2-case}.
}
\end{figure}

\begin{figure}[ht]
{\setlength{\unitlength}{1cm}
\begin{picture}(15.5,15.8)(-.7,0)
\psfrag{SQRTS}[l][l]{$\scriptstyle \sqrts\;[\gev]$}
\psfrag{SIGMA}[l][l]{$\scriptstyle \sigma\;[\fb]$}
\put(.1,6.5){$\tb = 6$}
\put(2.5,0.1){$m_{A^0} [\gev]$}
\put(5.2,7.4){$m_{H^0} [\gev]$}
\put(-5.5,-2){\resizebox{1.1\width}{1.1\height}{
\includegraphics*{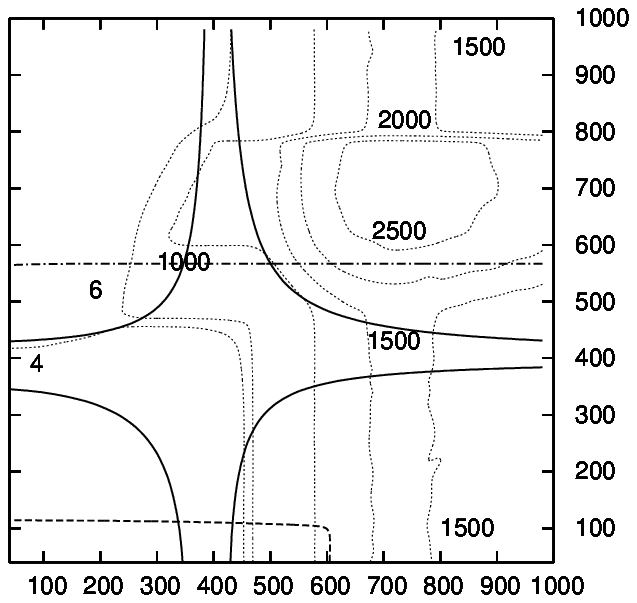}}}
\put(8.1,6.5){$\tb = 10$}
\put(10.5,0.1){$m_{A^0} [\gev]$}
\put(13.2,7.4){$m_{H^0} [\gev]$}
\put(2.5,-2){\resizebox{1.1\width}{1.1\height}{
\includegraphics*{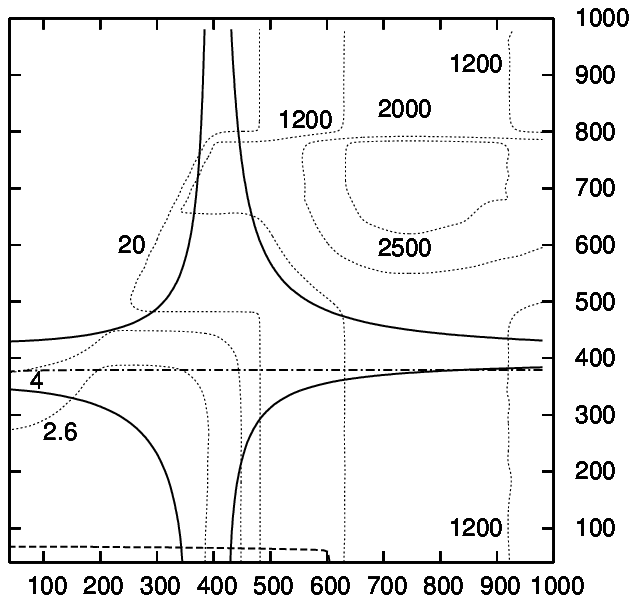}}}
\put(0.1,14.5){$\tb = 1.5$}
\put(10.5,8.1){$m_{A^0} [\gev]$}
\put(13.2,15.4){$m_{H^0} [\gev]$}
\put(-5.5,6){\resizebox{1.1\width}{1.1\height}{
\includegraphics*{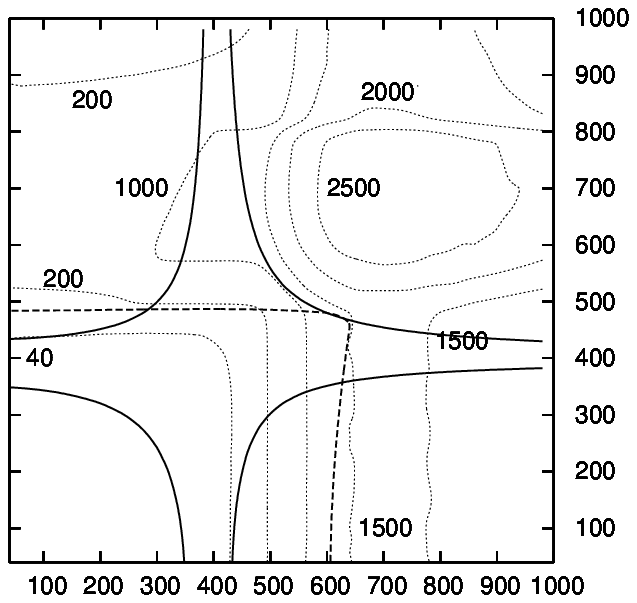}}}
\put(8.1,14.5){$\tb = 3$}
\put(2.5,8.1){$m_{A^0} [\gev]$}
\put(5.2,15.4){$m_{H^0} [\gev]$}
\put(2.5,6){\resizebox{1.1\width}{1.1\height}{
\includegraphics*{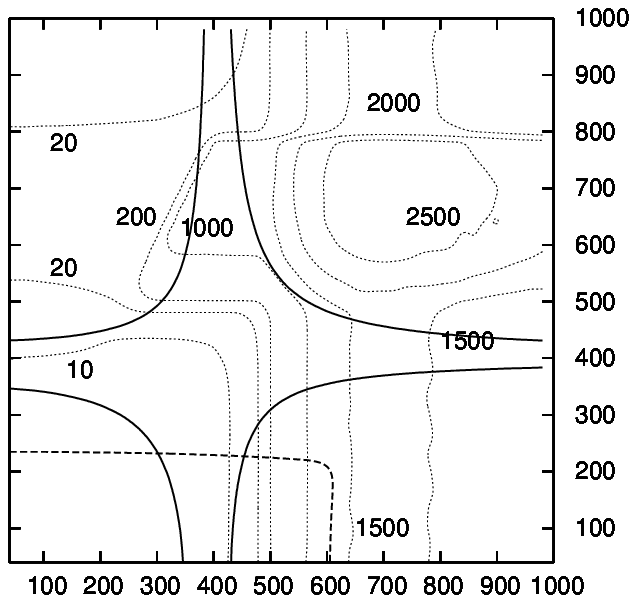}}}
\end{picture}
}
\caption{\label{m2-zero-case}
Contour of the Hadronic cross section $\sigma_{pp\to W^\pm H^\mp}$ 
in fb for the case $M^2=0$
(thin dotted lines). 
The limiting contours for the allowed values of
$\delta\rho_0$ (thick solid lines) allows the cross-shaped area, 
for $\delta a_\mu$ the whole displayed 
area for $\tb=1.5$, $3$ and the area below the thick dot-dashed line 
for $\tb=6$, $10$.
The perturbative unitarity constraint, 
$\text{max}\,|a_i| < 1/2$, (thick dashed lines) allows 
the almost rectangular area which includes the lower left corner.
The vacuum stability condition (\ref{eq:VS}) gives
no constraint on the displayed area.
}
\end{figure}

The hadronic cross section $\sigma_{pp\to W^\pm H^\mp}$ as a function of 
$\ma$ and $\mHH$, 
for $\tb = 1.5$, $3$, $6$, $10$
is displayed for $M^2 = m_{A^0}^2$ in Fig. \ref{m2-ma2-case}, 
$M^2 = m_{A^0}^2/2$ in Fig. \ref{m2-halfma2-case}, 
and $M^2 = 0$  in Fig. \ref{m2-zero-case}. 
In all Figures the exclusion limits (thick lines) 
are superimposed on the cross section contours (thin lines).
As a generic feature,
the cross section rises strongly,
if 
$\ma$ or $\mHH$ becomes larger than
$m_W+\mhpm \approx 480\,\gev$, 
with a maximum well above the contour 
where resonant $A^0$- and $H^0$-exchange contributions 
are possible.
This behavior 
is due to
two enhancement effects: 
a) resonant propagator contributions for $\ma,\mHH > 480\,\gev$ 
and 
b) a reduction of the strong negative interference if one 
propagator starts to decouple from the production threshold.
The variation of the cross section with $M$ 
is due to the $M$-dependence of the Higgs self-coupling constants 
(see Appendix \ref{tripleHcouplings}), 
which enter the widths of $A^0$ and $H^0$.

Imposing the constraints described in the previous chapter,
we focus on the remaining areas in Figs. \ref{m2-ma2-case} to
\ref{m2-zero-case} which are allowed by all constraints.
Remarkably, there remains an allowed region in all cases.
For $M^2=\ma^2$ (see Fig. \ref{m2-ma2-case}) the cross section in the allowed region
varies roughly in the range 20 to $1200\,\fb$ for $\tb=1.5$,
10 to $50\,\fb$ for $\tb=3$ and around the MSSM value 
for $\tb=6$ and $10$.
For $M^2=\ma^2/2$ (see Fig. \ref{m2-halfma2-case}) the lower bound 
of the cross section is similar to the previous case, while 
1000$\,\fb$ are possible for $\tb= 1.5$, 3 and 6, and 20$\,\fb$ for 
$\tb=10$, which is still about 60 times the MSSM cross section.
Depending on $\tb$, the case $M^2=0$ is strongly constrained to rather low 
values of $\mHH$. Because of the rho-parameter constraint, the allowed
areas for $\tb = 3, 6$  and 10
have a cross section below the corresponding MSSM scenario 
but of the same order.
For $\tb = 1.5$ again large variations of the cross section 
are possible, roughly between 40 and 1500$\,\fb$.

\section{Conclusions}
We have discussed the $H^\pm W^\mp$ production at hadron colliders 
in the framework of a general type II THDM
and compared the predictions with the MSSM.
We find that the THDM prediction for the hadronic cross section
can be completely different from the MSSM prediction.
Specifically, we find regions in parameter space where 
the cross section exceeds $1000\,\fb$.
These regions of large cross section are in agreement with
theoretical constraints, vacuum stability and perturbative unitarity,
and are not excluded by experimental constraints from measurements
of the rho-parameter and the muon magnetic moment.
In Ref.~\cite{mo}, it is discussed that
the size of the MSSM cross section would not be
sufficient to detect this process 
at the LHC.
An enhanced cross section in a general THDM scenario,
examples of which are shown in this paper, would give
a nice possibility to see the $H^\pm W^\mp$ signal at the LHC.
Certainly, the $H^\pm W^\mp$ cross section is not a 
discovery channel for the
charged Higgs at hadron colliders. However, once a charged Higgs boson
has been discovered, its observation will help to gain information
on the underlying model of the Higgs sector.
One should keep in mind that
there are many other new physics models,
apart from supersymmetry, which can be described by 
a THDM as a low energy effective theory.
It would be valuable to study the detectability of
$H^\pm W^\mp$ production in the framework of the general THDM 
by realistic simulation.
We provide a FORTRAN code for general use.

\bigskip\bigskip
\noindent {\em Acknowledgments.} 
S. K. would like to thank for the hospitality and support
during his stay in Aachen, where this paper was completed.
E. A. was supported by the Japan Society for the Promotion of Science.
S. K. was also supported in part by Grant-in-Aid of the Ministry of Education,
Culture, Sports, Science and Technology, Government of Japan, 
grant No. 17043008.
This work is supported in part by the Deutsche Forschungsgemeinschaft
Sonderforschungsbereich/Transregio 9, Computergest\"{u}tzte Theoretische Teilchenphysik.

\newpage

\appendix

\section*{Appendix}

\section{Higgs self-couplings}\label{tripleHcouplings}
We display here only those triple-Higgs couplings which are needed 
in the evaluation of the corresponding partial decay widths of the 
neutral Higgs bosons.
The case of a MSSM-like Higgs sector is recovered if the MSSM values 
are chosen for $m_{h^0}, m_{H^0}, \alpha$ and
$M$ is set to $\ma$.
The shorthands $s_\psi := \sin\psi, c_\psi := \cos\psi$ are used.
\begin{align}
\nonumber
 g_{h^0H^0H^0}  & =
        \sqrt{G_F}\bigg\{ 
	\frac{M^2}{\sqrt{2}}   
	(  
	s_{\alpha-\beta}
	- 3 \frac{s_\alpha}{c_\beta}
	+ 3 s_\alpha^2 \frac{c_{\alpha-\beta}}{c_\beta s_\beta}
)
       +   (4 m_{H^0}^2+ 2 m_{h^0}^2) 
	( \frac{s_\alpha}{c_\beta} 
	- s_\alpha^2 \frac{c_{\alpha-\beta}}{c_\beta s_\beta})
	\bigg\},\\
\nonumber
g_{h^0A^0A^0}  & =
        \sqrt{G_F}\bigg\{
	 \frac{M^2}{\sqrt{2}}   
	\frac{c_{\alpha+\beta}}{c_\beta s_\beta}
       + 4 m_{A^0}^2 
	s_{\alpha-\beta}
       - 2 m_{h^0}^2   
	(  
	\frac{c_{\alpha+\beta}}{c_\beta s_\beta}
	+ s_{\alpha-\beta})
	\bigg\},\\
\label{tripleHformulas}
g_{h^0H^+H^-}  & =
        \sqrt{G_F}\bigg\{
	 \frac{M^2}{\sqrt{2}}   
	\frac{c_{\alpha+\beta}}{c_\beta s_\beta}
       + 4 \mhpm^2 
	s_{\alpha-\beta}
       - 2 m_{h^0}^2   
	(  
	\frac{c_{\alpha+\beta}}{c_\beta s_\beta}
	+ s_{\alpha-\beta}
	)
	\bigg\},\\
\nonumber
 g_{H^0h^0h^0} & =
        \sqrt{G_F}\bigg\{
	 \frac{M^2}{\sqrt{2}}   
	( 
	- s_{\alpha+\beta}
	+ 3 \frac{s_\alpha}{s_\beta} 
	- 3 s_\alpha^2 \frac{s_{\alpha-\beta}}{s_\beta c_\beta}
	 )
       -   (4 m_{h^0}^2+ 2 m_{H^0}^2) 
	(  \frac{s_\alpha}{s_\beta} 
	- s_\alpha^2 \frac{s_{\alpha-\beta}}{s_\beta c_\beta}	
	)
	\bigg\},\\
\nonumber
g_{H^0A^0A^0} & =
        \sqrt{G_F}\bigg\{
	 \frac{M^2}{\sqrt{2}}   
	\frac{s_{\alpha+\beta}}{c_\beta s_\beta}
       - 4 m_{A^0}^2   
	c_{\alpha-\beta}
       +  2 m_{H^0}^2   
	( c_{\alpha-\beta} - \frac{s_{\alpha+\beta}}{c_\beta s_\beta} )
	\bigg\},\\
\nonumber
g_{H^0H^+H^-}  & =
        \sqrt{G_F}\bigg\{
	 \frac{M^2}{\sqrt{2}}   
        \frac{s_{\alpha+\beta}}{c_\beta s_\beta}
       - 4 \mhpm^2   
        c_{\alpha-\beta}
       + 2 m_{H^0}^2   
        ( c_{\alpha-\beta} - \frac{s_{\alpha+\beta}}{c_\beta s_\beta} )
	\bigg\} \;.
\end{align}

\section{Higgs and gauge boson scattering amplitudes}\label{s-wave-amp}
The tree-level S-wave amplitudes for the scattering of Higgs 
and longitudinal gauge bosons in the high energy regime can be calculated 
elegantly using the equivalence theorem \cite{PU1,equiv-theorem}.
We consider 14 distinct neutral channel 
scattering processes to be studied \cite{PU2},
\begin{align*}
W_L^+ W_L^-, W_L^+ H^-, W_L^- H^+, H^+ H^-, Z_L Z_L, Z_L A, A A, Z_L h, Z_L H, 
A h, A H, h h, h H, H H,
\end{align*}
which lead to a 14-dimensional S-matrix, with the following eigenvalues,
generically referred to as $a_i (i=1,\ldots,14)$,
expressed in terms of the parameters $\lambda_i$ 
of the Higgs potential (\ref{2hdm-higgs-pot}).
\begin{align}
\nonumber
a_\pm & =  \frac{1}{16 \pi} \Big\{  3 (\lambda_1 + \lambda_2 + 2 \lambda_3)
       \pm ( \sqrt{9 (\lambda_1-\lambda_2)^2 
		+ (4 \lambda_3+\lambda_4+\lambda_5/2+\lambda_6/2)^2}
          \Big\}\;,\\
\nonumber
b_\pm & = \frac{1}{16 \pi} \Big\{ \lambda_1+ \lambda_2+ 2 \lambda_3
       \pm \sqrt{(\lambda_1-\lambda_2)^2 
	+ (-2 \lambda_4+\lambda_5+\lambda_6)^2/4}
          \Big\}\;,\\
\nonumber
c_\pm & = d_\pm = \frac{1}{16 \pi}   \Big\{ \lambda_1+ \lambda_2+ 2 \lambda_3
       \pm \sqrt{(\lambda_1-\lambda_2)^2 + (\lambda_5-\lambda_6)^2/4}
          \Big\}\;,\\
\label{swave-amps}
e_1 & = \frac{1}{16 \pi} \Big\{2 \lambda_3 - \lambda_4 -\lambda_5/2 + 5 \lambda_6/2 \Big\}\;,\\
\nonumber
e_2 & = \frac{1}{16 \pi} \Big\{2 \lambda_3 + \lambda_4 -\lambda_5/2 +\lambda_6/2\Big\}\;,\\
\nonumber
f_+ & =  \frac{1}{16 \pi} \Big\{2 \lambda_3 - \lambda_4 + 5 \lambda_5/2 - \lambda_6/2\Big\}\;,\\
\nonumber
f_- & =  \frac{1}{16 \pi} \Big\{2 \lambda_3 + \lambda_4 + \lambda_5/2 - \lambda_6/2\Big\}\;,\\
\nonumber
f_1 & = f_2 = \frac{1}{16 \pi} \Big\{2 \lambda_3 + \lambda_5/2 + \lambda_6/2\Big\}.
\end{align}
In our study, we require the moduli of all these amplitudes 
to stay below 1/2~\cite{HHG}.

To obtain the S-wave amplitudes (\ref{swave-amps}) 
in terms of masses and mixing angles in the Higgs sector, 
we eliminate $\lambda_i (i=1,2,3,4,6)$ according to
the following relations.
\begin{align}
\nonumber
\lambda_1 & = \frac{\Sigma}{8v^2c_{\beta}^2} 
	+ \frac{\Delta}{8v^2}
		\Big(\frac{c_{2\alpha}}{c_{\beta}^2} - 2 \frac{s_{2\alpha}}{s_{2\beta}}\Big) 
	+ \frac{\lambda_5}{4} \Big(1 - \tan^2\beta\Big)\;,\\
\lambda_2 & = \frac{\Sigma}{8v^2c_{\beta}^2} 
	- \frac{\Delta}{8v^2}
		\Big(\frac{c_{2\alpha}}{s_{\beta}^2} + 2 \frac{s_{2\alpha}}{s_{2\beta}}\Big) 
	+ \frac{\lambda_5}{4} \Big(1 - \cot^2\beta\Big)\;,\\
\nonumber
\lambda_3 & = \frac{\Delta}{4v^2}\frac{s_{2\alpha}}{s_{2\beta}} 
	- \frac{\lambda_5}{4}\;,\;\;\;
	\lambda_4 = \frac{\mhpm^2}{v^2}\;,\;\;\;
	\lambda_6 = \frac{m_{A^0}^2}{v^2}\;,
\end{align}
with $\Sigma = m_{H^0}^2 + m_{h^0}^2$ and $\Delta = m_{H^0}^2 - m_{h^0}^2$.


\end{document}